\documentclass[structabstract,letter]{aa}  
\usepackage{graphicx}
\usepackage{txfonts}
\usepackage{natbib}
\usepackage{breqn}
\usepackage{longtable,lscape}
\newcommand{\BT}{\mbox{$B_T$}}
\newcommand{\VT}{\mbox{$V_T$}}
\newcommand{\GG}{\mbox{$G$}}
\begin{document}

   \title{Calibration of the photometric \GG\ passband for {\em Gaia} Data Release 1}


   \author{J. Ma{\'\i}z Apell{\'a}niz\inst{1}
          }

   \institute{Centro de Astrobiolog{\'\i}a, CSIC-INTA, campus ESAC, camino bajo del castillo s/n, E-28\,692 Villanueva de la Ca\~nada, Spain \\
              \email{jmaiz@cab.inta-csic.es} \\
             }

   \date{Received 25 October 2017; accepted 29 November 2017}

 
  \abstract
  {On September 2016 the first data from \mbox{{\em Gaia}} were released (DR1). The first release included photometry for over $10^9$ sources in the very broad \GG\ 
   system.}
  {To test the correspondence between \GG\ magnitudes in DR1 and the synthetic equivalents derived using spectral energy distributions from 
   observed and model spectrophotometry. To correct the \GG\ passband curve and to measure the zero point in the Vega system.}
  {I have computed the synthetic \GG\ and Tycho-2 \BT\VT\ photometry for a sample of stars using the Next Generation Spectral Library (NGSL) and the Hubble Space 
   Telescope (HST) CALSPEC spectroscopic standards.}
  {I have found that the nominal \GG\ passband curve is too blue for the DR1 photometry, as shown by the presence of a color term in the comparison 
   between observed and synthetic magnitudes. A correction to the passband applying a power law in $\lambda$ with an exponent of 0.783 eliminates the color term. The 
   corrected passband has a Vega zero point of $0.070\pm0.004$ magnitudes.}
  {}


   \maketitle
%

\section{Introduction}

\begin{figure}
\centerline{\includegraphics[width=0.9\linewidth]{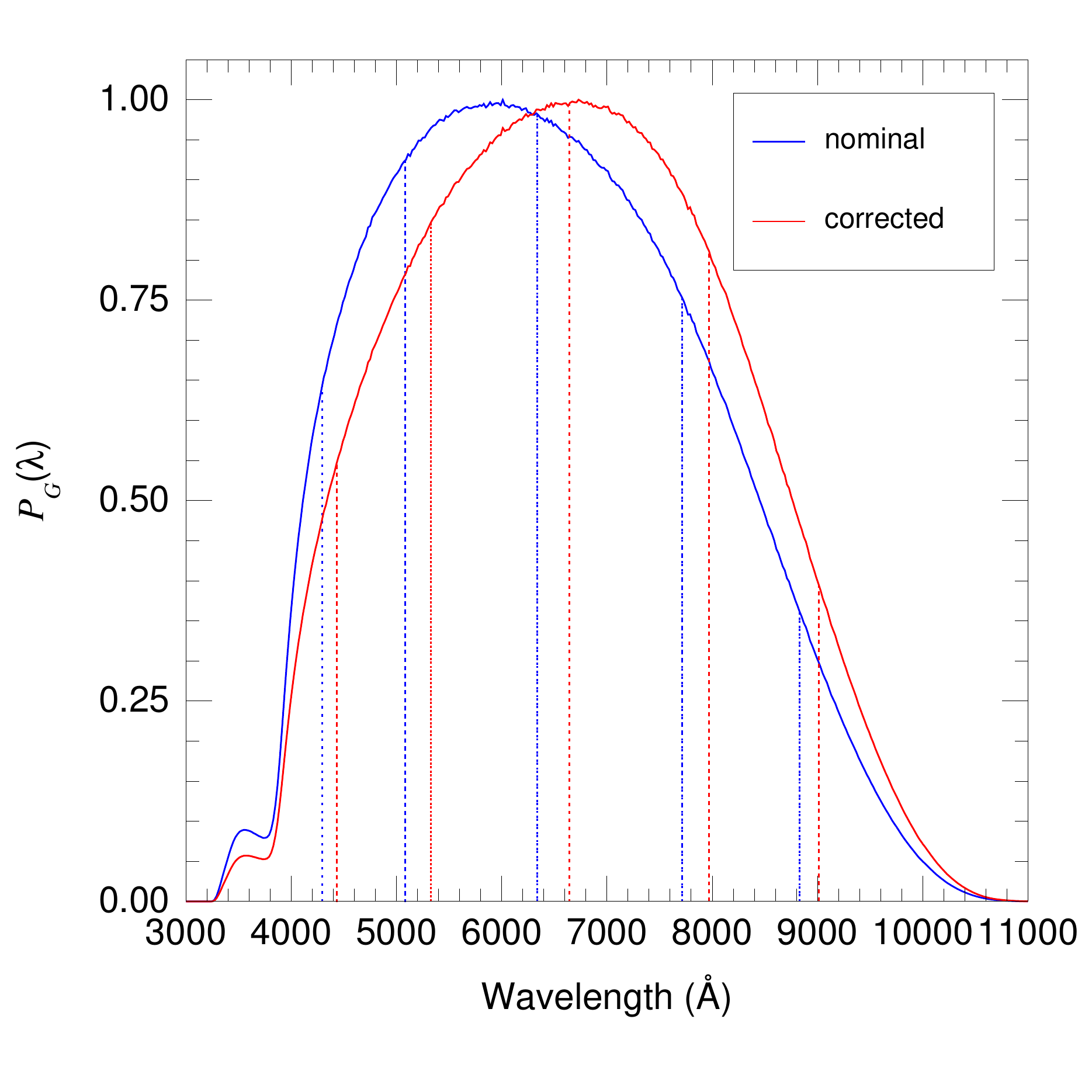}}
\caption{Nominal and corrected \GG\ passbands $P_G(\lambda)$. The dotted vertical lines mark the median and the $\pm\sigma$ and $\pm2\sigma$ gaussian percentile 
         equivalents for both passbands.}
\label{Gpassband}
\end{figure}

$\,\!$ \indent The first {\em Gaia} data release (DR1) was published in September 2016 and included positions and \GG\ magnitudes for over $10^9$ sources to 
\GG\ = 21 \citep{Browetal16}. The \GG\ passband is very broad, covering from the $U$ to the $y$ bands, even though the sensitivity at both extremes is rather low
(\citealt{Jordetal10} and Fig.~\ref{Gpassband}). The complex photometric calibration of such a large number of sources is presented by \citet{Carretal16} and 
involves both internal and external processes. 

\citet{Carretal16} described the existence of a color term in the external calibration of the \GG\ magnitude scale in the sense that, for a fixed observed value, a 
blue source was actually brighter than a red one. The difference between two unextinguished O and M stars amounts to about 0.2 magnitudes (see Fig.~14 in
\citealt{Carretal16}) and it arises because the nominal \GG\ passband of \citet{Jordetal10} differs from the true one. 
It is not uncommon to have small differences (sometimes of unknown origin) between a lab-measured passband and one measured once the instrument is operating.
In this case there was a known contamination effect caused by 
water freezing in some optical elements, a problem that affected the mission in its early observing stages \citep{Prusetal16}. \citet{Carretal16} 
indicated that the effect
would be solved in future data releases by publishing a modified \GG\ passband and suggested that, in the meantime, a color correction be applied to the observed
\GG\ magnitudes when comparing them with synthetic photometry. That strategy has two problems:

\begin{itemize}
 \item The color correction is a function of $G_{\rm BP}-G_{\rm RP}$, magnitudes that are not currently accessible as they will not be available until
       at least the second {\em Gaia} data release (DR2). Furthermore, the correction itself is not explicitly listed as it only appears in a plot. 
 \item Such color terms may be useful in some cases \citep{vanLetal17} but they cannot be applied in a general-purpose code for comparing observed and synthetic 
       photometry such as CHORIZOS \citep{Maiz04c}, which uses arbitrary filter sets and treats extinction in a detailed manner (extinguishing the spectral energy 
       distributions or SEDs and integrating over the band a posteriori).
\end{itemize}

\begin{figure*}
\centerline{\includegraphics[width=0.49\linewidth]{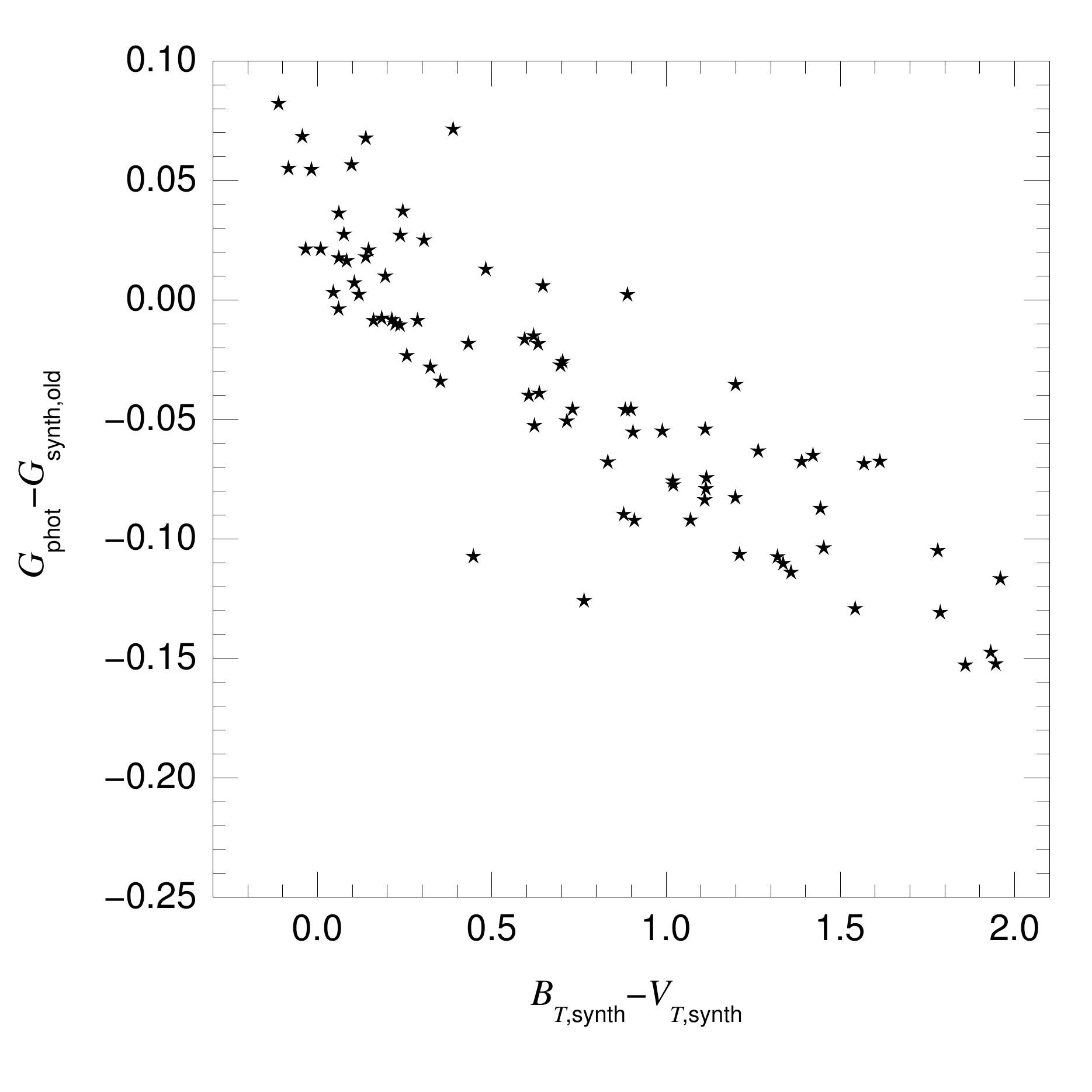} \
            \includegraphics[width=0.49\linewidth]{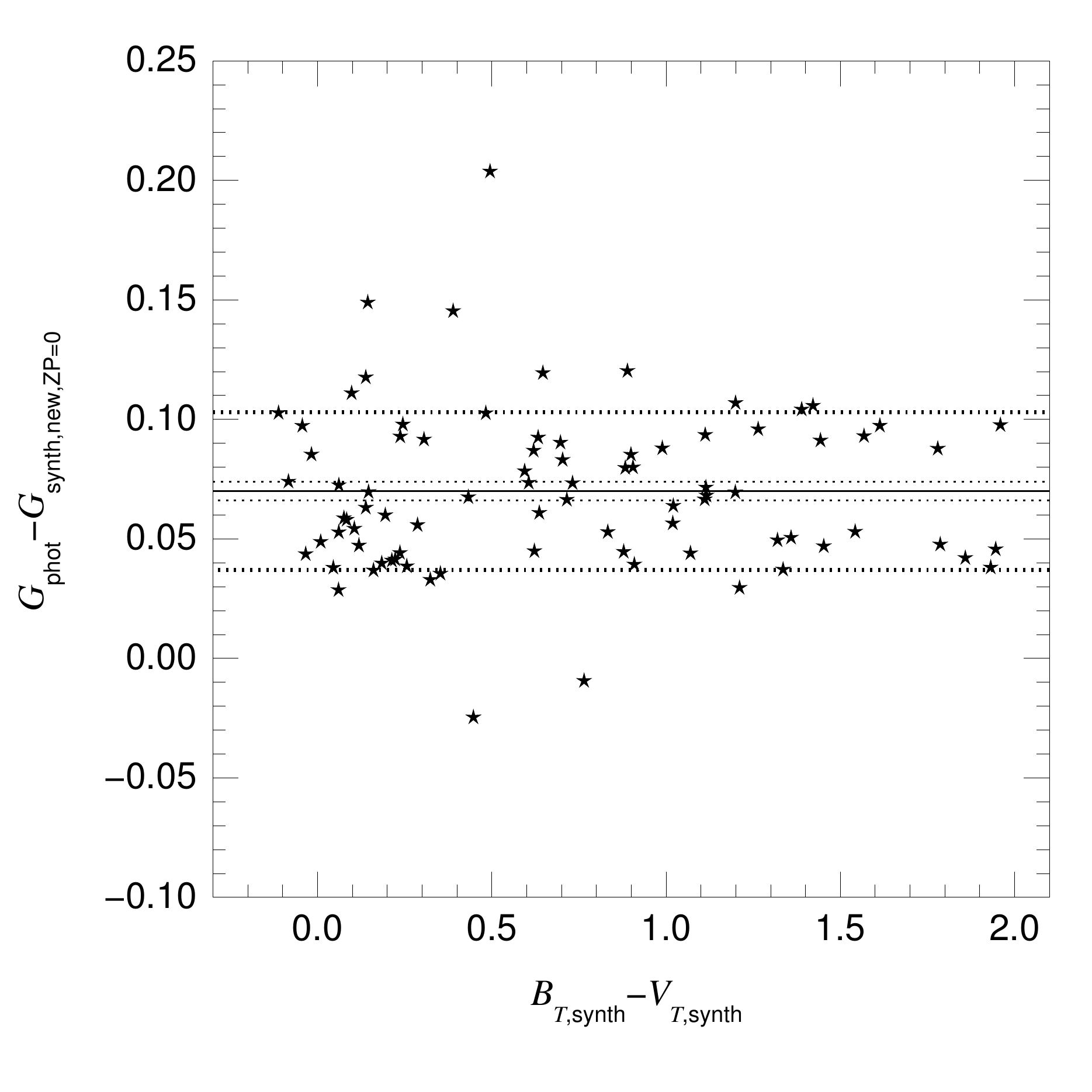}}
\caption{Difference between the photometric DR1 and the synthetic \GG\ magnitudes for the main NGSL+CALSPEC sample in this paper as a function of the 
         synthetic \BT$-$\VT\ color. The left panel uses the nominal \GG\ passband curve and Vega zero point (0.030 mag) while the right panel uses the corrected 
         \GG\ passband curve and a Vega zero point of 0. 
         The horizontal lines in the right panel show the mean value (solid line, the new zero point) and the range spanned 
         by the standard deviation of the data (thick dashed line, the typical uncertainty of an unsaturated star) and by the standard deviation of the mean (thin dashed line,
         the uncertainty of the zero point).
         }
\label{BVG}
\end{figure*}

For those reasons I decided to attempt a different approach: using existing information to generate a correction to the \GG\ passband, a technique I used
successfully in \citet{Maiz06a} for Johnson $U$ and Str\"omgren $u$. In that way, it should be possible to compare DR1 and synthetic photometry at this time without using 
color corrections and without having to wait for a future {\em Gaia} data release. That is the purpose of this letter.

\section{Results}

$\,\!$ \indent I define the \GG\ synthetic magnitude \citep{Maiz05b,Maiz06a,Maiz07a} as:

\begin{equation}
G_{\rm synth} = -2.5\log_{10}\left(\frac{\int P_G(\lambda)\,f_{\lambda}(\lambda)\,\lambda\,d\lambda}
                                        {\int P_G(\lambda)\,f_{\lambda{\rm,Vega}}(\lambda)\,\lambda\,d\lambda}\right)
                                        + {\rm ZP}_G,
\label{Gmag}
\end{equation}

\noindent where $P_G(\lambda)$ is the total-system passband or efficiency, $f_{\lambda}(\lambda)$ is the SED to be measured, $f_{\lambda{\rm,Vega}}(\lambda)$ is the 
Vega (reference) SED, and ${\rm ZP}_G$ is the zero point in the Vega system. Note that our $P_G(\lambda)$ includes three terms in the equivalent definition of 
\citet{Jordetal10}: $T$, $P$, and $Q$, but not $\lambda$ i.e. it is a photon-counting passband, not an energy-counting one \citep{Maiz06a}. The Vega SED is that of 
the CALSPEC file {\tt alpha\_lyr\_stis\_003.fits} \citep{Bohl07}. The nominal passband is that of \citet{Jordetal10}, also listed in Table~\ref{Gtable} here, with 
${\rm ZP}_G = 0.030$ mag.

\begin{table}
\caption{This table is available in electronic format at the CDS and it contains three columns: wavelength (in nm), the nominal total-system passband from 
         \citet{Jordetal10}, and the corrected one from this letter.}
\label{Gtable}
\end{table}

\begin{table}
 \caption{This table is available in electronic format at the CDS and it contains five columns: star name, Gaia DR1 ID, J2000 right ascension, J2000 declination, and a flag
          indicating the SED source (N for NGSL, C for CALSPEC).}
\label{Stable}
\end{table}

To test the validity of the nominal passband for Gaia DR1 magnitudes I collected STIS spectrophotometry from two sources: the Next Generation Spectral Library (NGSL,
\citealt{HeapLind07}) and the CALSPEC spectroscopic standards \citep{Bohletal17}. From those two sources I selected the stars with (a) accurate Tycho-2 \BT\VT\ and Gaia \GG\
photometry, eliminating objects with Tycho-2 variability flags and objects brighter than \GG\ = 6, where saturation starts taking place (see below for additional information
on \GG\ saturation) and (b) coverage of at least the 3000-10\,200~\AA\ range. The CALSPEC data were observed with a wide slit and require no further additional flux calibration.
The NGSL data were reduced with the same techniques used in \citet{Maiz05b} but with the additional step of recalibrating in flux using the \VT\ magnitude and the zero point
of \citet{Maiz07a}. As the \GG\ band has some sensitivity beyond 10\,200~\AA, the flux was extended until 11\,000~\AA\ using both a simple power law extension and 2MASS $J$ 
photometry (both alternatives yielded very small differences in $G_{\rm synth}$, at the level of 0.001 mag or less which, as will be shown, is significantly smaller 
than the effect that is being measured). A total of 84 stars satisfied the requirements and they are given in Table~\ref{Stable}.

The left panel of Fig.~\ref{BVG} shows the difference between the observed \GG\ magnitudes ($G_{\rm phot}$) and $G_{\rm synth}$ for the 84 stars in our sample as a function of 
the synthetic $\BT-\VT$ color using the nominal passband and zero point. There is a clear linear color term in the vertical scale that amounts to $\sim$0.2 magnitudes in the 2 
magnitudes range spanned by the $\BT-\VT$ color and that corresponds approximately to the difference between unextinguished O and M stars. The effect is consistent in sign and 
amplitude with the one found by \citet{Carretal16} discussed above, even though the samples and data used to measure it are different. Therefore, I confirm that the nominal 
\GG\ passband does not accurately describe the Gaia DR1 photometry and needs to be corrected, as already indicated by \citet{Carretal16}.

The likely source of the observed discrepancy (the presence of frozen water in some optical elements during the early stages of the mission) suggests that the passband should
be modified not by one or more discrete absorption bands but rather by multiplying it by a continuous smooth function. I chose as such a function a power law in $\lambda$,
i.e. a correction of the type $\lambda^\alpha$, and I iteratively tested different values of the exponent $\alpha$ to see the effect on $G_{\rm phot}-G_{\rm synth}$
in order to eliminate the color term in the left panel of Fig.~\ref{BVG}. After several attempts, the algorithm converged onto $\alpha = 0.783$, with lower values yielding
color terms with a negative slope and larger ones overcorrecting to produce a positive slope.
The result is shown in the right panel of Fig.~\ref{BVG}: the linear color term in $\BT-\VT$ has disappeared (a linear fit yields a slope that is essentially zero) and there
are no obvious second- or third-order terms. Therefore, I conclude that modifying the nominal \GG\ passband multiplying it by a power law with an exponent of 0.783 provides
a much improved characterization of the photometry when comparing it to observed spectrophotometry. The corrected \GG\ passband is shown in Fig.~\ref{Gpassband} and listed
in Table~\ref{Gtable}.

The right panel of Fig.~\ref{BVG} assumes ZP$_G$ of 0, as that value is a priori unknown. However, it can be easily calculated as the mean value of the data in the plot, 
which is 0.070. The plot also shows the standard deviation of the data, which is 0.033 mag. 
Some of that scatter is caused by the photometric uncertainties of the CALSPEC+NGSL data and once I remove that effect we are left with a dispersion of 0.030 mag.
That should be the value used to estimate the photometric uncertainty when 
comparing Gaia DR1 photometry with spectrophotometric models. Note that uncertainty is significantly greater than the published flux uncertainties in DR1. I suspect that this
is a consequence of the time-variable nature of the contamination, with different stars being observed at different points during the early stages of the mission. If that is 
the case, the Gaia DR2 should be more precise, as more epochs would have been included and the contamination effect would have a lower weight. The uncertainty on the zero
point itself can be estimated from the standard deviation of the mean and is 0.004 mag.

\begin{figure}
\centerline{\includegraphics[width=\linewidth]{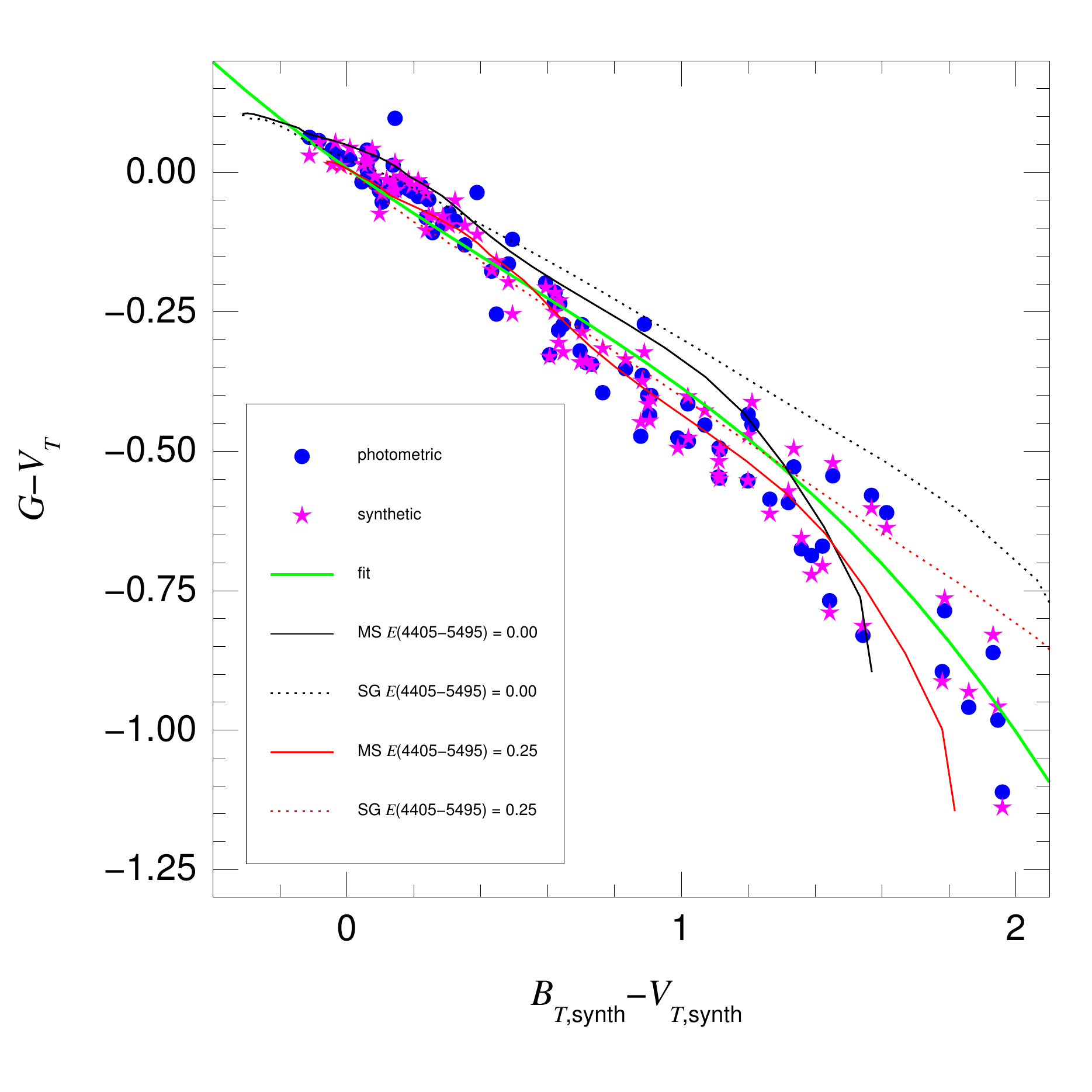}}
 \caption{\GG$-$\VT\ colors as a function of the synthetic \BT$-$\VT\ color. Both the observed (or photometric) and the synthetic (using the corrected \GG\ passband and the new 
          zero point) values are shown for the vertical axis. The green line is the cubic polynomial fit of \citet{vanLetal17}. The other four lines show synthetic photometry for
          main sequence (MS) and supergiant (SG) solar-metallicity 4000-40\,000~K models with no extinction and with $E(4405-5495) = 0.25$ and $R_{5495} = 3.1$ from 
          \citet{Maiz13a} and \citet{Maizetal14a}.}
\label{BVGV}
\end{figure}

As a test of the validity of our calibration, I show in Fig.~\ref{BVGV} the observed and synthetic $\GG-\VT$ colors for the sample of 84 stars as a function of the synthetic
$\BT-\VT$ colors, along with the cubic polynomial fit of \citet{vanLetal17}. Both the observed and synthetic colors show a good correspondence with the fit (other than the 
existence of possible higher-order fluctuations). The standard deviation of $(\GG-\VT)_{\rm phot}-(\GG-\VT)_{\rm synth}$ is 0.032 i.e. nearly identical to that of 
$\GG_{\rm phot}-\GG_{\rm synth}$. I also show the synthetic photometry 
for those colors
computed from the \citet{Maiz13a} grid with solar metallicity. The photometry is reproduced using
a wide range of $T_{\rm eff}$ and luminosities and requires a small degree of extinction in some of the stars, as expected from the NGSL+CALSPEC sample.

\begin{figure}
\centerline{\includegraphics[width=\linewidth]{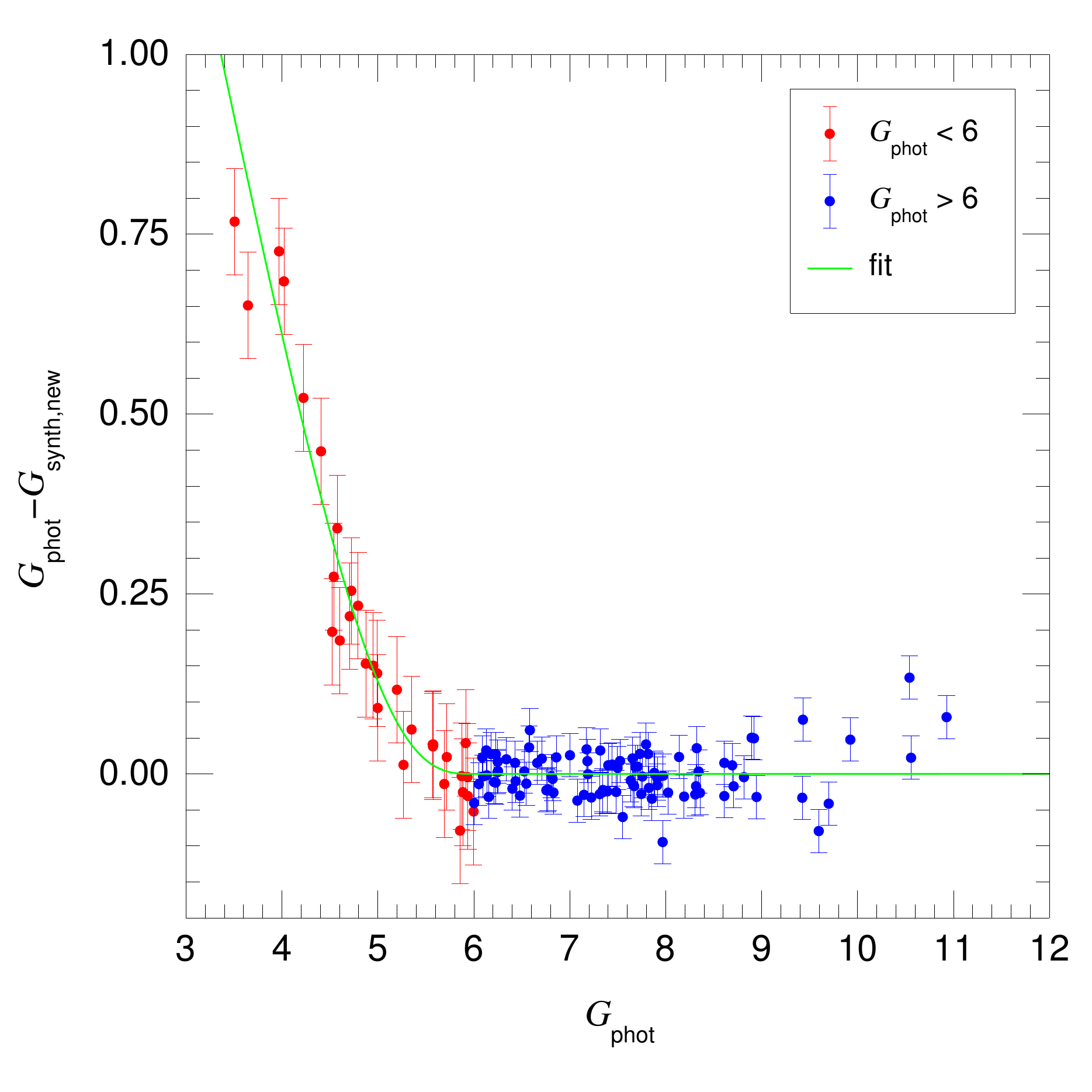}}
\caption{Difference between the observed DR1 and the synthetic magnitudes for the extended NGSL+CALSPEC sample in this paper as a function of the 
         observed \GG\ using the corrected \GG\ passband curve and the new zero point of 0.070 mag. Blue points are used for unsaturated objects ($\GG > 6$) and 
         red for saturated/near-saturated ones ($\GG < 6$). The green line shows the fit that can be used to correct for saturation in DR1 \GG\ magnitudes. The size of
         the blue error bars corresponds to the typical uncertainty of an unsaturated star (0.030 mag, Fig.~\ref{BVG}) while the red error bars correspond to the
         typical uncertainty of a saturated star, as determined from the standard deviation of the difference between the red points and the green fit (0.074 mag).}
\label{GG}
\end{figure}

An application of the new passband is the possibility of correcting saturated DR1 \GG\ magnitudes, which are expected to start at \GG = 6. I have selected an extended sample 
by relaxing the selection criteria to include NGSL stars brighter than that value, which yields an additional 33 stars in the range $\GG = 3.5-6.0$. The difference between the
observed and synthetic magnitudes as a function of the observed \GG\ is plotted in Fig.~\ref{GG}, where the onset of saturation is indeed shown to be close to $\GG = 6$ mag. Note,
however, that the scatter in the brighter stars is small. This led me to fit a saturation correction of the form:

\begin{equation}
\Delta = \frac{a\,(\GG-6)^3}{b+(\GG-6)^2}
\label{satcor}
\end{equation}

\noindent for $\GG < 6$ and 0 for $\GG \ge 6$, where $a$ and $b$ are two parameters to be fitted. The best fit yields $a = -0.56$ and $b = 3.39$ and the standard deviation between 
the data and the fit is 0.074 mag. Therefore, if one uses Eqn.~\ref{satcor} with those parameters to correct for \GG\ saturation, 0.074 mag should be the photometric uncertainty of
the resulting value. Note that the correction has not been tested for objects brighter than $G_{\rm phot} = 3.5$. Another application of the new passband is the combination of 
\GG\ magnitudes with other sources of photometry to calculate the extinction towards Galactic O stars (Ma\'{\i}z Apell\'aniz \& Barb\'a 2018, submitted to A\&A).

\section{Conclusions}

$\,\!$ \indent I have verified that the nominal \GG\ passband curve requires a correction for Gaia DR1 data, obtained such a correction, tested it sucessfully, and applied it to 
correct saturated magnitudes.

\begin{acknowledgements}
This work has made use of data from the European Space Agency (ESA) mission {\it Gaia} ({\tt https://www.cosmos.esa.int/gaia}), processed by the {\it Gaia} Data Processing and 
Analysis Consortium (DPAC, {\tt https://www.cosmos.esa.int/web/gaia/dpac/consortium}). Funding for the DPAC has been provided by national institutions, in particular the institutions 
participating in the {\it Gaia} Multilateral Agreement. I thank the anonymous referee for a constructive revision of the text and 
Carme Jordi for useful discussions on this topic.
I acknowledge support from the Spanish Government Ministerio de Econom{\'\i}a y Competitividad (MINECO) through grant AYA2016-75\,931-C2-2-P.
\end{acknowledgements}

\bibliographystyle{aa}
\bibliography{general}

\end{document}